\documentclass[aps,pra,onecolumn,notitlepage,11pt,superscriptaddress,floatfix,letterpaper,longbibliography,tightenlines]{revtex4-1}
\usepackage{amssymb,amsfonts,amsmath}

\usepackage{float} 

\usepackage{hyperref}
\hypersetup{colorlinks=true}
\usepackage[all]{hypcap} 

\usepackage{times}

\usepackage{upgreek}

\usepackage{graphicx}
\usepackage{color}

\newcommand{\ket}[1]{| #1 \rangle}

\newcommand{\ignore}[1]{}

\DeclareMathOperator{\re}{Re}
\DeclareMathOperator{\im}{Im}

\DeclareMathOperator{\Tr}{Tr}
\newcommand{\eq}{Eq.\,}
\newcommand{\eqs}{Eqs.\,}
\newcommand{\fig}{Fig.\,}

\newcommand{\cf} {cf.~}

\newcommand{\eg} {e.g.~}
\newcommand{\rref} {Ref.\,}
\newcommand{\rrefs} {Refs.\,}
\newcommand{\figurepanel}[2]{\hyperref[#1]{\ref*{#1}(#2)}}

\allowdisplaybreaks[1] 

\usepackage{filecontents}

\begin{document}
\title{Non-Markovian Dynamics of a Qubit Due to Single-Photon Scattering in a Waveguide}
\author{Yao-Lung L. Fang}
\altaffiliation[Present address: ]{Computational Science Initiative, Brookhaven National Laboratory, Upton, NY 11973-5000, USA}
\affiliation{Department of Physics, Duke University, P.O. Box 90305, Durham, North Carolina 27708-0305, USA}

\author{Francesco Ciccarello}
\affiliation{NEST, Istituto Nanoscienze-CNR and Dipartimento di Fisica e Chimica, Universita' degli Studi di Palermo, Italy}
\affiliation{Department of Physics, Duke University, P.O. Box 90305, Durham, North Carolina 27708-0305, USA}

\author{Harold U.\ Baranger}
\affiliation{Department of Physics, Duke University, P.O. Box 90305, Durham, North Carolina 27708-0305, USA}

\date{February 3, 2018}

\begin{abstract}
We investigate the open dynamics of a qubit due to scattering of a single photon in an infinite or semi-infinite waveguide. Through an exact solution of the time-dependent multi-photon scattering problem, we find the qubit's dynamical map. Tools of open quantum systems theory allow us then to show the general features of this map, find the corresponding non-Linbladian master equation, and assess in a rigorous way its non-Markovian nature. The qubit dynamics has distinctive features that, in particular, do not occur in emission processes. Two fundamental sources of non-Markovianity are present: the finite width of the photon wavepacket and the time delay for propagation between the qubit and the end of the semi-infinite waveguide.  
\end{abstract}
\maketitle


\section{Introduction}

\begin{figure}[b]
	\centering
	\includegraphics[scale=1.1]{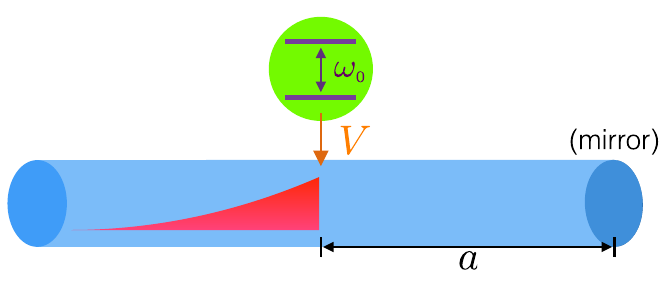}
	\caption{A qubit coupled to a waveguide scatters a single-photon wavepacket injected from the left. The qubit may start in an arbitrary state. The semi-infinite waveguide sketched here is terminated by an effective mirror, thus introducing a delay time.}
	\label{fig1}
\end{figure}

Waveguide quantum electrodynamics (QED) is an emerging area of quantum optics that investigates coherent coupling between one or more emitters (qubits) and a one-dimensional (1D) photonic waveguide \cite{LodahlRMP15,NohRPP16,LiaoPhyScr16,RoyRMP17,GuarXiv17}. Novel correlations among injected near-resonant photons result from the nonlinearity of the qubits, and intriguing interference effects occur because of the 1D confinement of the light. The field has focused on qubits in a local region for which these correlation and interference effects can be used for local quantum information purposes such as single-photon routing \cite{HoiPRL11}, rectification of photonic signals \cite{DaiPRA15,FratiniPRA16,MascarenhasPRA16,FangPRA17}, and quantum gates \cite{KoshinoPRA10,CiccarelloPRA12,ZhengPRL13a}. This regime of waveguide QED involves neglecting \emph{delay times}: the time taken by photons to travel between qubits is far shorter than all other characteristic times. However, an important goal for photonic waveguides is to carry out \emph{long-distance} quantum information tasks such as quantum state transfer between remote quantum memories \cite{CiracPRL97,KimbleNat08}. As these necessarily involve distant qubits, delay times cannot be neglected, leading to different kinds of photon correlation and interference effects through the \emph{non-Markovian} (NM) nature of the system. Here, we study a model waveguide-QED system with large delay time.  
We apply recent developments in the theory of open quantum systems (OQS) 
in order to quantitatively assess the qubit's degree of non-Markovianity. 

A large variety of waveguide-QED setups have been experimentally demonstrated in recent years \cite{RoyRMP17,GuarXiv17,LauchtPRX12,GobanNatComm14}. Because of high photon group velocities and small systems, these experiments are mostly described by a Markovian approach in which delay times are neglected. In contrast, recent experiments have started entering the regime of non-negligible delay times \cite{RochPRL14,GustafssonSci14,SundaresanPRX15,ChantasriPRX16,LiuNatPhys16}, an area that is expected to grow rapidly due to interest in extended systems and long-distance quantum information. Accounting for photon delay times is, however, a challenging theoretical task:
only recently have the dynamical effects of long delay times started being investigated \cite{TufarelliPRA13,GonzalezBallestroNJP13,ZhengPRL13,TufarelliPRA14,RedchenkoPRA14,LaaksoPRL14,FangPRA15,GrimsmoPRL15,ShiPRA15,RamosPRA16, GuimondQST17}. 
A major consequence of delay times is that NM effects are introduced that affect the physics profoundly, as predicted for \eg qubit-qubit entanglement in emission \cite{GonzalezBallestroNJP13,RamosPRA16} and second-order correlation functions in photon scattering \cite{ZhengPRL13,LaaksoPRL14,FangPRA15,ShiPRA15}.

Clarifying the importance of NM effects and the mechanisms behind their onset is thus pivotal in waveguide QED. 
At the same time, the theory of OQS \cite{BreuerBook,RivasHuelgaBook} is currently making major advances, yielding a more accurate understanding of NM effects \cite{BreuerJoPB12, RivasRPP14,BreuerRMP16,deVegaRMP17}. Through an approach often inspired by quantum information concepts \cite{Nielsen00}, a number of physical properties such as information back-flow \cite{BreuerPRL09} and divisibility \cite{RivasPRL10} have been spotlighted as distinctive manifestations of quantum NM behavior and then used to formulate corresponding quantum \emph{non-Markovianity measures}.
These tools have been effectively applied to dynamics in various scenarios \cite{RivasRPP14,BreuerRMP16}, including in waveguide QED with regard to \emph{emission} processes \cite{emission,RamosPRA16, TufarelliPRA14} such as a single atom emitting into a semi-infinite waveguide \cite{TufarelliPRA14}.

Motivated by the need to quantify NM effects in photon \emph{scattering}  from qubits, we present a case study of a qubit undergoing single-photon scattering in an infinite or semi-infinite waveguide (see \fig\ref{fig1}), the latter of which is the basis of the proposed controlled-\textsc{Z} \cite{CiccarelloPRA12} and controlled-\textsc{Not} \cite{ZhengPRL13a} gates. 
We aim at answering two main questions: \textit{What are the essential features of the qubit open dynamics during scattering? Is such dynamics NM?} 
The key task is to find the dynamical map (DM) of the qubit in the scattering process, which fully describes the open dynamics and is needed in order to apply OQS tools \cite{BreuerRMP16}. 
A distinctive feature of our open dynamics is that the bath (the waveguide field) is initially in a well-defined single-photon state \cite{MirzaPRA16a,ValenteOL16}.
Toward this task, we tackle in full the time evolution of \textit{multiple excitations} (in contrast to those limited to the one-excitation sector \cite{ChenNJP11,LiaoPRA15,LiaoPRA16,MirzaPRA16a,XiongPhyA17}), a problem that has become important recently \cite{LongoPRL10, LongoPRA11, PeropadrePRL13, 
ZuecoFD14, GrimsmoPRL15,ShiPRA15, KocabasPRA16, KocabasOL16, MirzaPRA16b, GuoPRA17, WhalenQST17,GuimondQST17}.

Intuitively, one may expect that the dynamics is fully Markovian in the infinite-waveguide case and NM in the semi-infinite case due to the atom-mirror delay time. We show that this expectation is inaccurate in general, 
mostly because it does not account for a fundamental source of NM behavior namely the wavepacket bandwidth. This NM mechanism is present in an infinite waveguide, 
while in a semi-infinite waveguide it augments the natural NM behavior coming from the photon delay time. Recently, NM effects in infinite-waveguide scattering were addressed in \rref\cite{ValenteOL16}. 
There, however, the qubit is always initially in the ground state, while a fair application of non-Markovianity measures should be based on the entire DM thus requiring consideration of an arbitrary initial state of the qubit.


The paper is organized as follows. We first define the system under consideration in Sec.\,\ref{sec: system}. Next, in Sec.\,\ref{sec: DM}, we find the general form of the the qubit's DM in a single-photon scattering process and discuss its main features. In Sec.\,\ref{sec: ME}, we present the time-dependent master equation (ME), which is fulfilled exactly by the qubit state at any time. In Sec.\,\ref{sec: explicit construction}, we discuss the explicit computation of the dynamical map in the infinite- and semi-infinite-waveguide case (most of the details regarding the former are given in the Appendix). Since this task requires the time evolution of the scattering process, we in particular find a closed delay partial differential equation that holds in the two-excitation sector of the Hilbert space for a semi-infinite waveguide. In Sec.\,\ref{sec: NM}, we assess the non-Markovian nature of the scattering DM by making use of non-Markovianity measures. In this way, we identify two fundamental sources of NM behavior: the finiteness of the wavepacket width and the time-delayed feedback due to the mirror. We finally draw our conclusions in Sec.\,\ref{sec: conclusion}. Some technical details are given in the Appendices.

\section{System}
\label{sec: system}
Consider a qubit with ground (excited) state $|g\rangle$ ($|e\rangle$) and frequency $\omega_0$, which is coupled at $x=x_0$ to a photonic waveguide (along the $x$-axis) with linear dispersion. We model the system via the standard \cite{ShenPRA07,ShenPRA09I,FangPRA15} real-space Hamiltonian under the rotating-wave approximation (we set $\hbar=c=1$ throughout) 
\begin{align}
	\hat H&=-i\int\limits_{-\infty}^\xi\! {\rm d}x\!\left[\hat a_\text{R}^\dagger(x)\partial_x \hat a_\text{R}(x)- \hat a_\text{L}^\dagger(x)\partial_x \hat a_\text{L}(x)\right]+\omega_0\hat \sigma_+\hat \sigma_- \nonumber\\
	&\quad+V\int\limits_{-\infty}^\xi \!{\rm d}x\,\delta(x-x_0)\left[\left(\hat a^\dagger_\text{R}(x)+ \hat a^\dagger_\text{L}(x)\right)\hat \sigma_-
	+{\rm h.c.}\right],
	\label{H}
\end{align}
where the bosonic operator $\hat a_\text{R}(x)$ [$\hat a_\text{L}(x)$] annihilates a right-going (left-going) photon at $x$, $\hat \sigma_-= \hat \sigma_+^\dag= |g\rangle\langle e|$, and $V$ is the qubit-field coupling strength such that the qubit decay rate into the waveguide is $\Gamma= 2V^2$. For an infinite waveguide, the upper integration limit is $\xi=+\infty$ and $x_0=0$, while for a semi-infinite waveguide $\xi=0^+$ and $x_0=-a$ (see \fig\ref{fig1}).

\section{Dynamical map}
\label{sec: DM}

By definition the qubit's DM, $\Phi_t$, is the superoperator that when applied on \emph{any} qubit state at $t=0$, $\rho_0$, returns its state at time $t>0$ \cite{BreuerBook,RivasHuelgaBook}, 
\begin{equation}
\rho(t)=\Phi_t [\rho_0].\label{DM}
\end{equation}
The DM fully specifies the open dynamics of the qubit coupled to the waveguide field, with the latter serving as the reservoir. 

We now find the DM for a single-photon wavepacket. 
Let $\hat U_t = e^{-i \hat H t}$
be the unitary evolution operator of the joint qubit-field system. The initial state for single-photon scattering is  
$\sigma_0= \rho_0\,|\varphi\rangle\langle \varphi|$ (tensor product symbols are omitted),
where $\rho_0 = \rho_{gg}|g\rangle\langle g|+\rho_{ee}|e\rangle\langle e|+\left(\rho_{ge}|g\rangle\langle e|+{\rm h.c.}\right)$ and $|\varphi\rangle = \int \!{\rm d}x \,\varphi(x) \,\hat a_{\rm R}^\dag(x)|{0}\rangle$
is the incoming single-photon (normalized) wavepacket ($|{0}\rangle$ is the waveguide vacuum state).
At time $t$, the atom-field state is $\sigma(t)=\hat U_t \,\sigma_0\hat U^\dag_t =\hat U_t \rho_0\,|\varphi\rangle\langle \varphi|  \hat U^\dag_t$. 
By plugging $\rho_0$ into $\sigma(t)$,
we get
\begin{equation}
\sigma(t) =  \rho_{gg}\hat U_t |g\varphi\rangle\langle g\varphi|\hat U^\dag_t 
+\rho_{ee}\hat U_t |e\varphi\rangle\langle e\varphi|\hat U^\dag_t 
+\left[\rho_{ge}\hat U_t |g\varphi\rangle\langle e\varphi|\hat U^\dag_t +{\rm H.c.}\right],
\label{sigmat2}
\end{equation}
hence for any $\rho_0$ the knowledge of the pair of elementary unitary processes $\hat U_t |g\varphi\rangle$ and $\hat U_t |e\varphi\rangle$ fully specifies the time evolution of $\sigma(t)$. Due to the conservation of the total number of excitations [see \eqref{H}], the joint evolved state in the two processes has the form
\begin{align}
|\Psi_1(t)\rangle&=\hat U_t |g\rangle|\varphi\rangle=|g\rangle\,|\phi_1(t)\rangle+e(t)\, |e\rangle|{0}\rangle,\label{ugphi}\\
|\Psi_2(t)\rangle&=\hat U_t |e\rangle|\varphi\rangle=|g\rangle\,|\chi_2(t)\rangle+|e\rangle\,|\psi_1(t)\rangle,\label{uephi}
\end{align}
where $|\Psi_n(t)\rangle$ is the joint wavefunction at time $t$ for $n$  excitations. Here, $|\phi_1(t)\rangle$ and $|\psi_1(t)\rangle$ are unnormalized single-photon states, and $|\chi_2(t)\rangle$ is an unnormalized two-photon state. 
Note that (\ref{ugphi}) [(\ref{uephi})] describes the joint dynamics of a single photon scattering off a qubit initially in the ground [excited] state, which takes place entirely in the one-excitation [two-excitation] sector of the Hilbert space. In particular, Eq.~\eqref{uephi} is a \textit{stimulated emission} process \cite{RephaeliPRL12}.

The qubit state at time $t$ is the marginal $\rho(t)=\Tr_\text{field}\sigma(t)$. 
This partial trace can be performed by placing \eqs(\ref{ugphi}) and (\ref{uephi}) into \eq (\ref{sigmat2}), which yields
\begin{align}
\rho(t)
&=
\left[\rho_{gg}\Vert\phi_1(t)\Vert^2 + \rho_{ee}\Vert\chi_2(t)\Vert^2\right]|g\rangle\langle g|+
\left[\rho_{gg}|e(t)|^2 + \rho_{ee}\Vert\psi_1(t)\Vert^2\right]|e\rangle\langle e|\nonumber\\
&\quad+\left[\rho_{ge}\langle \psi_1(t)|\phi_1(t)\rangle|g\rangle\langle e|+{\rm H.c.}\right],
\label{rhot}
\end{align}
where we took advantage of orthogonality between one-photon and two-photon states. 
Since $\{|g\varphi\rangle$, $|e\varphi\rangle\}$ are normalized, so are (\ref{ugphi}) and (\ref{uephi}) due to unitarity of $\hat U_t$. Thus,
$\Vert\phi_1(t)\Vert^2+|e(t)|^2=\Vert\chi_2(t)\Vert^2+\Vert\psi_1(t)\Vert^2=1$. 
By defining three time functions
\begin{equation}
p_g(t)\equiv|e(t)|^2,\quad
p_e(t)\equiv\Vert\psi_1(t)\Vert^2,\quad
c(t)\equiv\langle\phi_1(t)|\psi_1(t)\rangle,
\label{3functions}
\end{equation}
we have $\Vert\phi_1(t)\Vert^2=1-p_g(t)$ and $\Vert\chi_2(t)\Vert^2= 1- p_e(t)$. Therefore, changing to the matrix representation, \eq (\ref{rhot}) takes the form (with $\rho_{ee}=1- \rho_{gg}$)
\begin{equation}
\Phi_t[\rho_0]=\left(
\begin{array}{cc}
p_e(t)- \Delta (t)\rho_{gg}&c(t) \rho_{eg}\\
c^*(t)\rho_{eg}^*  & 1- p_e(t)+ \Delta (t)\rho_{gg}
\end{array}
\right),
\label{phit1}
\end{equation}
where we defined 
\begin{equation}
	\Delta (t)\equiv p_e(t)- p_g(t).
	\label{eq: Delta}
\end{equation}
Note that \textit{both} $p_g(t)$ and $p_e(t)$ are the qubit excited-state populations but in the two \textit{different} processes (\ref{ugphi}) and (\ref{uephi}), 
respectively.

We refer to the qubit DM \eqref{phit1}
as the ``scattering DM.'' Since the atom-field initial state $\sigma_0$ is a product state and $\hat U_t$ is unitary, the map $\Phi_t$ is necessarily completely positive (CP) and trace preserving \cite{Nielsen00}. In contrast to pure emission processes at zero temperature \cite{BreuerBook}, here both the one- and two-excitation sectors are involved.
We stress that the DM is fully independent of the qubit's initial state $\rho_0$, being dependent solely on the Hamiltonian (\ref{H}) and the field initial state $|\varphi\rangle$. This dependance occurs through the functions of time $p_{g/e}(t)$ and $ c(t)$ in (\ref{phit1}), where $p_{g/e}(t)$ determine the qubit populations while $c(t) $ governs the coherence. 

The DM's form is best understood in the Bloch-sphere picture \cite{Nielsen00} in which a qubit state $\rho$ is represented by the Bloch vector 
${\bf r}=(2\re \rho_{ge}, 2\im\rho_{ge}, \rho_{gg}-\rho_{ee})$ with $\Vert{\bf r}\Vert\le1$. In this picture, the map $\Phi_t$ is defined by the vector identity 
\begin{equation}
{\bf r}(t)={\bf M}_t {\bf r}_0+{\bf v}(t),
\label{rt}
\end{equation}
where ${\bf v}(t)=\left(0,0,1-p_e(t)-p_g(t)\right)^{\rm T}$ and ${\bf M}_t$ is the $3\times3$ matrix 
\begin{equation}
{\bf M}_t=\!\left(\begin{array}{cc}
|c(t)| {\bf R}_{\theta(t)} & {\bf 0}^{\rm T}\\
{\bf 0} & \Delta (t)
\end{array}\right).
\label{Mt}
\end{equation}
Here, ${\bf 0}=(0,0)$ while ${\bf R}_{\theta(t)}$ is a standard $2\times2$ rotation matrix of angle $\theta(t)=\arg[c(t)]$.
Thus, apart from the rigid displacement ${\bf v}(t)$ and rotation around the $Z$-axis, the scattering process shrinks the magnitude of the $XY$- and $Z$-components of ${\bf r}(t)$ by the factors $ \left|c(t)\right|^2$ and $\left|\Delta(t)\right|$, respectively. Since these two factors are generally unequal, the DM transforms the Bloch sphere into an \emph{ellipsoid}. Such a lack of spherical symmetry does not occur in emission processes \cite{LorenzoPRA13}, thus providing a hallmark of  scattering open dynamics. 

In addition, a careful look at \eqs(\ref{rt}) and (\ref{Mt}) shows that 
the Bloch vector undergoes a reflection across the $XY$-plane whenever $\Delta(t)<0$. This is a further distinctive trait of scattering dynamics, not occurring in emission processes \cite{LorenzoPRA13}, which is shown below to be relevant to the onset of NM behavior.

\section{Non-Markovian Master Equation}
\label{sec: ME}

The most paradigmatic Markovian dynamics is the one described by the celebrated Lindblad ME \cite{BreuerBook}, 
\begin{equation}
\dot\rho=-i [\hat{\cal H},\rho]+ \sum_\nu\gamma_\nu{\cal L}_\nu[\rho], 
\;\;\textrm{with}\;\; {\cal L}_{\nu}[\rho]=\hat L_\nu\rho\hat L_\nu^\dag-(\hat L_\nu^\dag\hat L_\nu\rho+\rho\hat L_\nu^\dag\hat L_\nu) /2 ,
\end{equation} 
where $\hat{\cal H}$ is self-adjoint, and all the rates $\gamma_\nu$'s are \emph{positive constants}. 
In our case, the DM (\ref{phit1}) is not described by a Lindblad ME;
instead, we show that it is described by a time-dependent ME \cite{brazil}:
\begin{equation}
   \dot\rho = -i [\hat{\mathcal{H}}(t),\rho] 
   +\gamma_+(t)\mathcal{L}_+[\rho] 
   +\gamma_-(t)\mathcal{L}_-[\rho] 
   +\gamma_z(t)\mathcal{L}_z[\rho] 
   \label{ME}
\end{equation}
where $\hat{\mathcal{H}}(t)$ is a time-dependent Hamiltonian, and the jump operators $\hat{L}_{\nu}=\hat \sigma_\nu$ describe three non-unitary channels with the \emph{time-dependent} rates $\gamma_+(t)$ for absorption, $\gamma_-(t)$ for emission, and $\gamma_z(t)$ for pure dephasing [explicit forms are given below in \eqs\eqref{eq: time-dep decay rates}].
We note that the dephasing term, which reflects the lack of spherical symmetry of the evolved Bloch sphere discussed above, does not occur in spontaneous emission. 

The general form for a time-dependent ME is
\begin{equation}
\dot\rho=\mathcal L_t[\rho],
\label{Lrho}
\end{equation}
where $\mathcal L_t$ is a time-dependent linear (and traceless) map, which is fulfilled by $\rho(t)$ as given by \eq(\ref{phit1}).
The standard recipe for carrying this out starting from the DM is to first take the derivative of \eq(\ref{DM}), which yields $\dot\rho=\dot\Phi_t[\rho_0]$. Introducing next the inverse of map $\Phi_t$, $\Phi_t^{-1}$, we can replace $\rho_0= \Phi_t^{-1}[\rho(t)]$. Hence, 
\begin{equation}
\mathcal L_t=\dot\Phi_t[\Phi_t]^{-1}.
\label{L2}
\end{equation}
The task now reduces to explicitly calculating $\mathcal L_t$ and expressing it in a Lindblad form so as to end up with \eq(\ref{ME}). 

This task is efficiently accomplished in the generalized 4-dimensional Bloch space. Recall that the set of four Hermitian operators $\{\hat G_i\}= \{\openone/\sqrt{2},\hat \sigma_x/\sqrt{2},\hat \sigma_y/\sqrt{2},\hat \sigma_z/\sqrt{2}\}$ --- where $i=0,1,2,3$, respectively --- is a basis in the qubit operators's space and fulfills $\Tr\{\hat G_i\hat G_j\}= \delta_{ij}$. 
We express both Eqs.\,(\ref{ME}) and (\ref{L2}) in this basis and equate them; some details are presented in Appendix \ref{sec: MEcalc}. The resulting expressions for the time-dependent Hamitonian as well as the three time-dependent rates in ME (\ref{ME}) are given by
\begin{subequations}
	\label{eq: time-dep decay rates}
	\begin{align}
	\hat{\mathcal{H}}(t)&=-\im \left[\frac{\dot c(t)}{c(t)} \right]\hat \sigma_+\hat \sigma_-\\
	\gamma_+(t)&=\frac{p_e(t)\dot p_g(t)- p_g(t) \dot p_e(t)}{ p_e(t)- p_g(t)},\\
	\gamma_-(t)&= -\frac{{ \dot p_e(t)- \dot p_g(t) }}{ p_e(t)- p_g(t)} -\gamma_+(t),\\
	\gamma_z(t)&= -\frac{\gamma_+(t)+ \gamma_-(t)}{4}- \re \frac{\dot c(t)}{2c(t)}.
	\end{align}
\end{subequations}
It can be checked that when $\hat{\mathcal{H}}(t)$ and these rates are placed in it, ME (\ref{ME}) is exactly fulfilled by Eq.\ (\ref{phit1}) at all times $t$.

Before concluding this section, we note that an exact, differential system (DS) governing the same open dynamics that applies in the case of an infinite waveguide was worked out in \rref\cite{GheriChapter05} and more recently further investigated and generalized in \rrefs\cite{BaragiolaPRA12,GoughPRA12}. For the present case of a single-photon wavepacket and a qubit, this DS has overall three unknowns: two density matrices, one of which is $\rho(t)$, and a traceless non-Hermitian matrix. In contrast to ME \eqref{ME} here, the DS has the advantage that its time-dependent coefficients are known functions of the wavepacket functional shape (in the time domain). However, since such a DS is not closed with respect to $\rho(t)$ it is less suitable for analyzing the general properties of the qubit's dynamical map, which is a major goal of the present work. Finally, while ME (\ref{ME}) holds for a qubit coupled to a generic bosonic bath under the rotating-wave approximation, 
the DS relies on the further hypothesis of a {\it white-noise} bosonic bath (hence, in particular, it does not hold in the semi-infinite-waveguide case).

\section{Explicit computation of DM}
\label{sec: explicit construction}

For the initial state of the waveguide, throughout this paper
we consider
an exponential incoming wavepacket of the form 
\begin{equation}
\varphi(x) =  i\sqrt{\alpha\Gamma}\exp{\left[(ik+\alpha\Gamma/2) (x-x_0)\right]}\theta(-x+x_0),
\label{eq: exponential wavepacket}
\end{equation}
where $k$ is an arbitrary central frequency,  $\alpha\equiv\delta k/\Gamma$ is the wavepacket bandwidth in units of 
$\Gamma$, and $\theta(x)$ is the step function. 
This choice of the wavepacket shape is often made (see \eg\cite{RephaeliPRL12}) as it has at least three advantages. An exponential shape allows for closed-form solutions in the Laplace domain in some cases. In addition, in a numerical approach, its well-defined wavefront leads to a significant reduction in computational time, which is important in the two-photon sector when there is a long time delay. Finally, such wavepackets can be generated experimentally \cite{HouckNat07,BozyigitNatPhys10,PierreAPL14,PechalPRX14,PengNatComm16,FronDiazPRApplied17} by either spontaneous emission of a qubit or tunable, on-demand sources. 

Our general approach is to plug the \emph{ansatz} for $\ket{\Psi_n(t)}$, \eqs(\ref{ugphi}) and (\ref{uephi}), into the Schr\"{o}dinger equation  $i\partial_t \ket{\Psi_n(t)}=\hat{H}\ket{\Psi_n(t)}$ to obtain a system of differential equations for the amplitudes that we solve for the three functions $p_g(t)$, $p_e(t)$, and $c(t)$ in (\ref{3functions}) and hence for the DM (\ref{phit1}). For an infinite waveguide, this can be accomplished analytically in both the one- and two-excitation sectors as shown in Appendix~\ref{sec: inf wg calculation}. Here, we  focus on the far more involved case of a semi-infinite waveguide. 

\begin{figure}[t]
	\centering
	\includegraphics[scale=0.7]{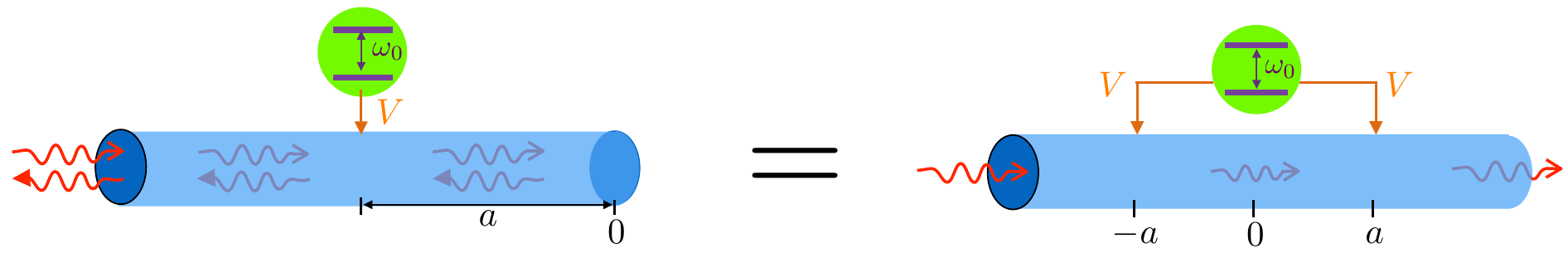}
	\caption{A qubit coupled to a semi-infinite waveguide with qubit-mirror distance $a$ (left) and its equivalent model featuring a qubit coupled to a \emph{chiral} infinite waveguide at the two points $x=\pm a$ (right).}
	\label{fig:semi-infinite waveguide unfolding}
\end{figure}

In this case, it is convenient to ``unfold'' the waveguide semi-axis at the mirror ($x=0$) by introducing a chiral field defined on the entire real axis 
by 
$\hat{a}(x) =\hat{a}_\text{R}(x)\theta(-x)-\hat{a}_\text{L}(x)\theta(x)$
(the minus sign encodes the $\pi$-phase shift due to reflection from the mirror);
see Fig.~\ref{fig:semi-infinite waveguide unfolding}.
The Hamiltonian \eqref{H} can then be rewritten 
by expressing $\hat{a}_{\text{R}/{\text L}}(x)$ in terms of $\hat{a}(x)$ as
\begin{equation}
\hat{H}=-i\int_{-\infty}^\infty \!dx\,\hat{a}^\dagger(x)\partial_x\hat{a}(x)
+\omega_0\hat{\sigma}_+\hat{\sigma}_-
-i V\!\int_{-\infty}^\infty \!dx\!\left[\delta(x+a)-\delta(x-a)\right]\left[\hat{a}^\dagger(x)\hat{\sigma}_- -\hat{\sigma}_+\hat{a}(x)\right].
\label{eq:Hamiltonian semi-infinite}
\end{equation}
Compared to \eq\eqref{H}, the form of the free-field term shows that only one propagation direction is allowed (chirality) while the term $\propto V$ shows the \emph{bi-local} coupling of the field to the qubit at points $x=\pm a$ (these can be seen as the locations of the real qubit and its mirror image, respectively).

\subsection{Semi-infinite waveguide: one-excitation sector}
The wavefunction ansatz in the one-excitation sector is [\cf\eq(\ref{ugphi})]
\begin{equation}
|\Psi_1(t)\rangle=\int dx\, \phi(x, t)\hat{a}^\dagger(x)|0\rangle+e(t)\hat\sigma_+|0\rangle,
\end{equation}
which once inserted into the Schr\"odinger equation yields the pair of coupled differential equations
\begin{subequations}
\begin{align}
	i\partial_t\phi(x,t)&=-i\partial_x\phi(x,t)-i V e(t)\left[\delta(x+a)-\delta(x-a)\right],\\
	i\frac{d}{dt}e(t)&=\omega_0 e(t)+iV\left[\phi(-a, t)-\phi(a, t)\right].
	\label{eq:single-excitation differential equation semi-infinite}
\end{align}
\end{subequations}
Integration of the photonic part yields the formal solution
\begin{equation}
	\phi(x, t)=\phi(x-t,0)-V\left[e(t-x-a)\theta(x+a)\theta(t-x-a)
	-e(t-x+a)\theta(x-a)\theta(t-x+a)\right],
	\label{eq:single-photon formal solution semi-infinite}
\end{equation}
[cf.~Eq.~(17) of Ref.~\cite{TufarelliPRA13}], which once plugged back into the equation for $e(t)$ [\eq(\ref{eq:single-excitation differential equation semi-infinite})] yields the delay (ordinary) differential equation (DDE)
\begin{equation}
	\frac{de(t)}{dt}=-\left(i\omega_0+\frac{\Gamma}{2}\right)e(t)+\frac{\Gamma}{2}e(t-2a)\theta(t-2a)+\sqrt{\frac{\Gamma}{2}}\left[\phi(-a-t,0)-\phi(a-t,0)\right].
	\label{eq:single-excitation delay differential eq appendix}
\end{equation}
Equation~\eqref{eq:single-excitation delay differential eq appendix} is the same as the well-known DDE describing the spontaneous emission process in Refs.~\cite{CookPRA87,DornerPRA02,TufarelliPRA13} but with the presence of the extra source term
$\propto\sqrt{\Gamma/2}$ due to the incoming photon wavepacket.
For the initial conditions $e(0)=0$ and $\phi(x,0)=\varphi(x)$ [\cf\eq(\ref{ugphi})], the solution for $e(t)$ obtained by Laplace transform reads
\begin{align}
	e(t)&=\frac{\sqrt{\alpha\Gamma^2/2} (e^{-(i\omega_0+\Gamma /2)t}-e^{-(ik+\alpha\Gamma/2)t})}{k-\omega_0+i\Gamma/2(1-\alpha)} 
	-i\sqrt{\alpha\Gamma} \sum_{n=1}^{\infty}\frac{\left(\frac{\Gamma}{2}\right)^{n-1/2}}{n!}
	\Bigl[(t-2na)^n e^{-(i\omega_0+\Gamma/2)(t-2na)}\nonumber \\
	&\quad\quad+\frac{i^n(k-\omega_0-i\alpha\Gamma/2)}{\left[k-\omega_0+i\Gamma/2(1-\alpha)\right]^{n+1}}\gamma(n+1,-ip(t-2na))e^{-(ik+\alpha\Gamma/2)(t-2na)}
	\Bigr]\theta(t-2na)
	\label{eq:exact solution single-photon exponential}
\end{align}
where $p=k-\omega_0+i\Gamma/2(1-\alpha)$ and $\gamma(n,z)$ is the incomplete Gamma function \cite{NISThandbook}. The corresponding solution for $\phi(x,t)$ follows straightforwardly by using \eqref{eq:exact solution single-photon exponential} in \eq(\ref{eq:single-photon formal solution semi-infinite}).

\subsection{Semi-infinite waveguide: two-excitation sector}

\begin{figure}[tb]
	\centering
	\includegraphics[scale=0.6]{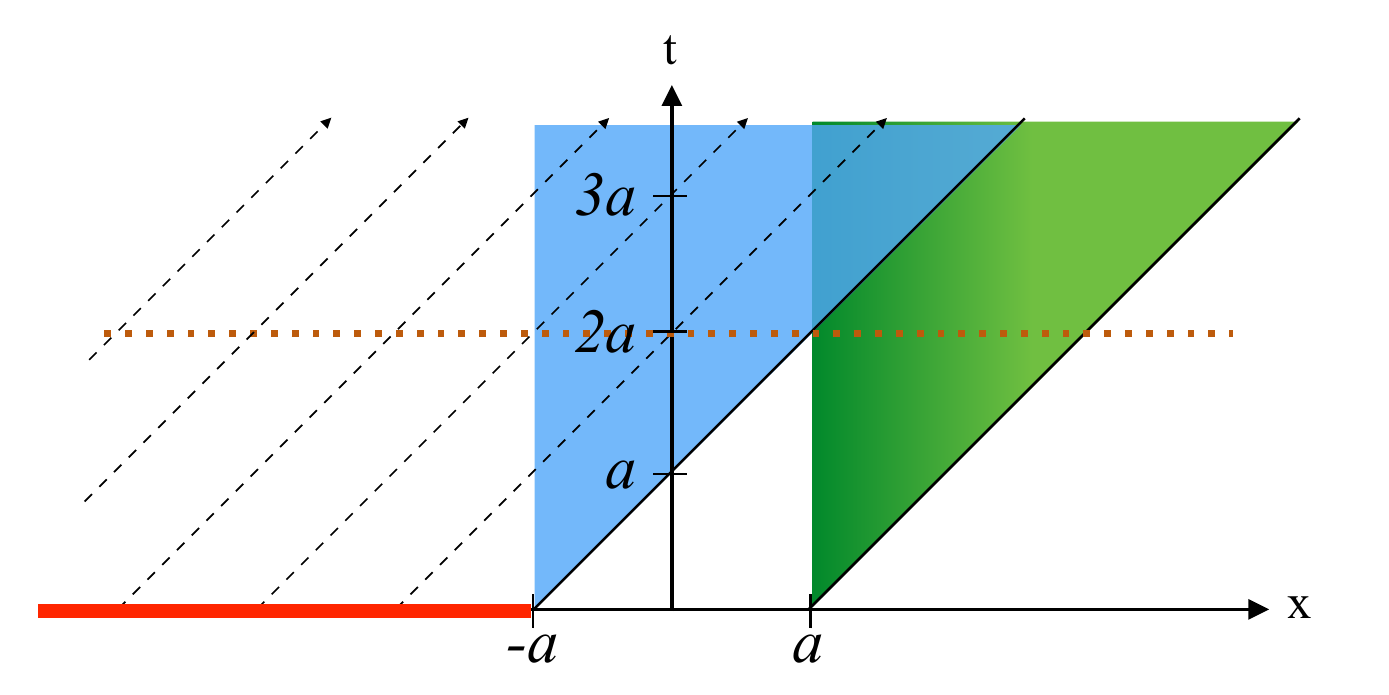}
	\caption{Spacetime diagram for Eq.~\eqref{eq:double-excitation delay differential eq full}. The dashed lines represent the propagation direction of the initial condition (thick red line). The brown dotted line shows the time after which the delay term appears. The blue and green regions are the light cones of the qubit and its mirror image, respectively.}
	\label{fig:spacetime-chiral}
\end{figure}

The ansatz for the time-dependent wavefunction [\cf\eq(\ref{uephi})] reads
\begin{equation}
	|\Psi_2(t)\rangle=\int dx\, \psi(x, t)\hat{a}^\dagger(x)\hat\sigma_+|0\rangle
	+\iint dx_1dx_2\,\chi(x_1, x_2, t)\frac{\hat{a}^\dagger(x_1)\hat{a}^\dagger(x_2)}{\sqrt{2}}|0\rangle.
	\label{eq:double-excitation wavefunction semi-infinite}
\end{equation}
The Schr\"odinger equation then yields the system of coupled differential equations
\begin{subequations}
	\begin{align}
		i\partial_t\psi(x, t)&=-i \partial_x\psi(x,t)+\omega_0\psi(x,t)+\frac{iV}{\sqrt{2}}\left[\chi(x,-a,t)+\chi(-a,x,t)-\chi(x,a,t)-\chi(a,x,t)\right]\label{dpsi},\\
		i\partial_t\chi(x_1,x_2,t)&=-i\left(\partial_{x_1}+\partial_{x_2}\right)\chi(x_1,x_2,t)-\frac{iV}{\sqrt{2}}\left[\psi(x_1,t)\left(\delta(x_2+a)-\delta(x_2-a)\right)+x_2 \leftrightarrow x_1\right].
	\end{align}
\end{subequations}
The formal solution for $\chi(x_1, x_2, t)$ is thus 
\begin{align}
	\chi(x_1, x_2, t)& = \chi(x_1-t, x_2-t, 0)-\frac{V}{\sqrt{2}}\biggl[
	\psi(x_1-x_2-a, t-x_2-a)\theta(x_2+a)\theta(t-x_2-a)\nonumber\\
	&\quad
	-\psi(x_1-x_2+a, t-x_2+a)\theta(x_2-a)\theta(t-x_2+a) +\bigl(x_2 \leftrightarrow x_1\bigr)
	\biggr],
	\label{eq:formal solution of two-photon wavefunction}
\end{align}
where note that $\chi$ is symmetrized under the exchange $x_1\leftrightarrow x_2$. By placing \eq(\ref{eq:formal solution of two-photon wavefunction}) into \eq(\ref{dpsi}) we find a 
spatially non-local delay partial differential equation (PDE) for $\psi(x,t)$:
\begin{equation}
\begin{aligned}
	\partial_t\psi(x,t)&=-\partial_x\psi(x,t)-\left(i\omega_0+\frac{\Gamma}{2}\right)\psi(x,t)+\frac{\Gamma}{2}\psi(x-2a, t-2a)\theta(t-2a)\\
	&\quad- \frac{\Gamma}{2}\biggl\{\bigl[\psi(- x- 2a, t- x- a) - \psi(- x, t- x- a)\bigr]\theta(x+ a)\theta(t- x- a)\\
	&\quad\quad+ \bigl[\psi(2a- x, t- x+ a)- \psi(- x, t- x+ a)\bigr]\theta(x- a)\theta(t- x+ a)\biggr\}\\
	&\quad+\sqrt{\frac{\Gamma}{4}}\bigl[\chi(x-t,-a-t,0)+\chi(-a-t,x-t,0)-\chi(x-t,a-t,0)-\chi(a-t,x-t,0)\bigr].
	\label{eq:double-excitation delay differential eq full}
\end{aligned}
\end{equation}
Equation~\eqref{eq:double-excitation delay differential eq full} is the two-excitation-sector counterpart of the DDE \eqref{eq:single-excitation delay differential eq appendix}. 
Mathematically, such a spatially non-local delay PDE is far more involved than the DDE \eqref{eq:single-excitation delay differential eq appendix} or conventional delay PDEs \cite{Zubik_Kowal_Sch08} that are local in space. A spacetime diagram is shown in Fig.~\ref{fig:spacetime-chiral}.

In our case [\cf\eq(\ref{uephi})], 
the initial conditions are $\psi(x,0)=\varphi(x)$ and $\chi(x_1, x_2, 0)=0$. Due to the latter condition, the terms on the last line of \eq(\ref{eq:double-excitation delay differential eq full}) are identically zero. Hence, overall, the differential equation features \emph{four source terms} that are non-local in $x$ and $t$ and are non-zero for $x>-a$. In the region $x\le-a$, the equation takes the simple form
\begin{equation}
	\partial_t\psi(x,t)=-\partial_x\psi(x,t)-\left(i\omega_0+\frac{\Gamma}{2}\right)\psi(x,t)+\frac{\Gamma}{2}\psi(x-2a, t-2a)\theta(t-2a).
	\label{eq:delay PDE simplified}
\end{equation}
By taking the Fourier (Laplace) transform with respect to variable $x$ ($t$), this equation is turned into an algebraic equation whose solution is given by
\begin{align}
	\bar{\psi}(q, s)
	&=\frac{\tilde{\varphi}(q)}{s+iq+(i\omega_0+\frac{\Gamma}{2})-\frac{\Gamma}{2}e^{-2a(s+iq)}}\nonumber\\
	&=\frac{\tilde{\varphi}(q)}{s+i(q+\omega_0)+\frac{\Gamma}{2}} 
	\sum_{n=0}^{\infty}\left[\frac{\frac{\Gamma}{2} e^{-2a(s+iq)}}
	{s+i(q+\omega_0)+\frac{\Gamma}{2}}\right]^n,
	\label{eq:general solution psi in q-s domain}
\end{align}
where $\tilde{\varphi}(q)={\sqrt{\frac{\alpha\Gamma}{2\pi}}e^{iqa}}/({k-q-i\frac{\alpha\Gamma}{2}})$ is the Fourier transform of $\varphi(x)$. Performing the inverse Fourier transform with respect to $q$ then yields
\begin{align}
	\tilde{\psi}(x,s)
	&=\frac{\sqrt{\alpha\Gamma}}{2\pi}\int dq\,\frac{e^{iq[x-(2n-1)a]}\sum\limits_{n=0}^{\infty}\left[\frac{\frac{\Gamma}{2} e^{-2as}}
		{s+i(q+\omega_0)+\frac{\Gamma}{2}}\right]^n }{(k-q-i\frac{\alpha\Gamma}{2})\left[s+i(q+\omega_0)+\frac{\Gamma}{2}\right]} \nonumber\\
	&=\frac{i\sqrt{\alpha\Gamma}e^{(ik+\alpha\Gamma/2)(x+a)}}{s+i(k+\omega_0)+\frac{\Gamma}{2}(1+\alpha)}\sum_{n=0}^{\infty}\left[\frac{\frac{\Gamma}{2} e^{-2a(s+ik+\alpha\Gamma/2)}}{s+i(k+\omega_0)+\frac{\Gamma}{2}(1+\alpha)}\right]^n,
	\label{eq: psi in x-s domain exponential}
\end{align}
where we used that, since $x<-a$, only the pole $q=k-i\alpha\Gamma/2$ contributes to the integral. Upon inverse Laplace transform with respect to $s$ term by term, we finally find
\begin{equation}
	\psi(x,t)=i\sqrt{\alpha\Gamma}e^{i(k-i\frac{\alpha\Gamma}{2})(x-t+a)}
	\left\{e^{-(i\omega_0+\frac{\Gamma}{2})t}\sum_{n=0}^{\infty}\frac{1}{n!}\left[\frac{\Gamma}{2}e^{(i\omega_0+\frac{\Gamma}{2})2a}(t-2na)\right]^n\theta(t-2na)\right\}
	\label{eq:exact solution two-photon exponential}
\end{equation}
for $x<-a$. 
This solution can be expressed compactly as $\psi(x,t) = \varphi(x-t)e_{\rm sm}(t)$, where $e_{\rm sm}(t)$ is the qubit excited-state amplitude in the spontaneous emission process \cite{CookPRA87,DornerPRA02,TufarelliPRA13} namely the solution of \eq(\ref{eq:single-excitation delay differential eq appendix}) for the initial conditions $e(0)=1$ and $\phi(x,0)=0$. This is physically clear: since the qubit starts in the excited state [\cf\eq(\ref{uephi})] so long as the photon has not reached its location $x=-a$ the system's evolution consists of the free propagation of the input single-photon wavepacket and the spontaneous emission as if the field were initially in the vacuum state.

The next natural step would be finding the wavefunction for $-a\le x \le a$. However, a look at Eq.~\eqref{eq:double-excitation delay differential eq full} shows that such a task is non-trivial. Specifically, two of the source terms, $\psi(-x-2a, t-x-a)$ and $-\psi(-x, t-x-a)$, enter the differential equation which forces one to find the solution ``tile by tile'' as discussed in the Supplementary Material \cite{SupMat}, a challenging and in the end impractical task. 
We choose instead to solve the delay PDE numerically by adapting the finite-difference-time-domain (FDTD) method \cite{FDTDbook,FDTDnote,FangFDTD17}; our approach is described in Ref.~\cite{FangFDTD17}.  
Note that the effectiveness of our code is crucially underpinned by the knowledge of the exact solution for $x<-a$ discussed above \cite{FangFDTD17}.

Finally, once the solutions in both number sectors are found, the three functions $p_{e/g}(t)$ and $c(t)$ can be obtained explicitly as 
\begin{equation}
	p_g(t) = |e(t)|^2,\quad p_e(t) = \int dx\,|\psi(x,t)|^2, \quad c(t) = \int dx\,\phi^*(x,t)\psi(x,t),
\end{equation}
where $e(t)$ is given by Eq.~\eqref{eq:exact solution single-photon exponential}, $\phi(x, t)$ is given by Eq.~\eqref{eq:single-photon formal solution semi-infinite}, and $\psi(x, t)$ is obtained from FDTD.

\section{Non-Markovianity}
\label{sec: NM}
Despite having a Lindblad structure, the time-dependent ME \eqref{ME} is not in general a Lindblad ME, not even one whose Lindblad generator is time-dependent,
because the rates $\{\gamma_\nu(t)\}$ are not necessarily \emph{all positive at all times} \cite{BreuerBook,RivasHuelgaBook,HallPRA14,BreuerRMP16}. The condition $\gamma_\nu(t)\ge 0$ for any $\nu$ and $t$ is indeed violated if $\Delta(t)<0$ at some time $t$ [recall the definition of $\Delta(t)$ in Eq.~\eqref{eq: Delta}].

Indeed, since $\Delta(0)=1$, if $\Delta$ becomes negative during the time evolution then there exists an instant at which both $\dot{\Delta}<0$ and $\Delta<0$. Then, since $\gamma_++\gamma_- = -\dot{ \Delta }/ \Delta$ [see \eq(\ref{ME}) and related text], at least one of the rates $\{\gamma_\pm(t)\}$ 
must be negative at some time. 
When this happens, the DM is not ``CP-divisible'': the dynamics cannot be decomposed into a sequence of infinitesimal CP maps \cite{BreuerBook} each associated with a time $0\le t'\le t$ and fulfilling (\ref{ME}) with positive rates $\{\gamma_{\nu}(t')\}$. Equivalently, it is not governed by a Lindblad ME even locally in time 
\cite{nota}. 
Thus, according to the criteria in \rrefs\cite{RivasPRL10, HallPRA14}, the dynamics is NM. Note that the negativity of $\Delta(t)$ is also sufficient to break P-divisibility (a weaker property than CP-divisibility) since it ensures that the sum of at least a pair of time-dependent rates in ME \eqref{ME} is negative \cite{BreuerRMP16}. 

\begin{figure}[tbp]
	\centering
	\includegraphics[scale=1.2]{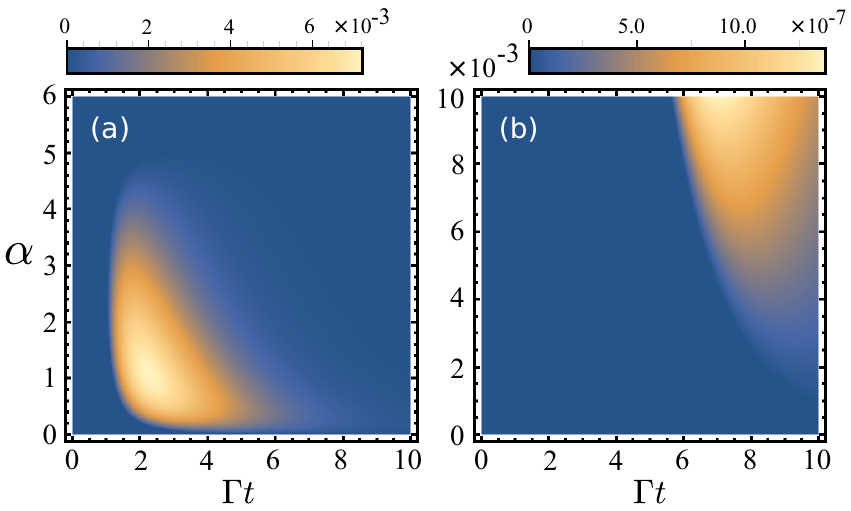}
	\caption{Density plot of the negativity of $\Delta$ 
		as a function of both time $t$ (in units of $\Gamma^{-1}$) and the wavepacket width $\alpha$ for an infinite waveguide. Panel (b) is a small-$\alpha$ zoom of (a). 
		($k=\omega_0= 20 \Gamma$.) 
		\label{fig2}}
\end{figure}

Negativity of $\Delta$ can occur already for an infinite waveguide 
if the wavepacket width is in an optimal range. Indeed, from the analytic expressions for $p_{e/g}(t)$ in the infinite-waveguide case given in Appendix~\ref{sec: 3 functions}, we get (time in units of $\Gamma^{-1}$ and $k=\omega_0$)
\begin{equation}
\Delta(t)=e^{-(\alpha +1) t}\,\frac{8 \alpha e^{\frac{(\alpha +1)t}{2}}+(\alpha -5) (\alpha +1) e^{\alpha  t}+4(1-\alpha)}{\alpha ^2-1}.
\label{lambdat2}
\end{equation}
Based on elementary analysis, two behaviors are possible (Appendix~\ref{sec: analysis of Delta}): for $\alpha>5$, $\Delta(t)\ge0$ always, while for $0<\alpha\le5$, $\Delta(t)$ has a minimum at a negative value at some time. 
To illustrate this, we plot $\Delta$'s negativity, $N_\Delta(t)\equiv-{\rm min}[0,\Delta(t)]$, in \fig\ref{fig2}.
For $0<\alpha\le5$, $N_\Delta (t)$ is initially zero, then exhibits a maximum at a time of the order of $\Gamma^{-1}$ and eventually decays to zero. For $\alpha\lesssim10^{-2}$, this maximum is in fact negligible reaching at most $N_\Delta\sim\! 10^{-6}$ [Fig.~\figurepanel{fig2}{b}]. For practical purposes, then, $\Delta(t)$ becomes negative for an optimal range of wavepacket widths $\delta k$ around $\alpha\simeq1$, i.e., $\delta k\simeq\Gamma$, which excludes small $\alpha$ and hence in particular quasi-plane waves.

Rigorously speaking, it should be noted that --- as typically happens with time-convolutionless MEs \cite{BreuerRMP16} ---  in general there may be singular times at which ME (\ref{ME}) is not defined and correspondingly the DM not invertible. In the present case, these are the times at which $\Delta(t)$ and/or $c(t)$ vanish [\cf\eqs\eqref{eq: Delta}, \eqref{eq: time-dep decay rates}, and \eqref{eq: p_g p_e c inf wg}]. The above sufficient condition should thus in general be complemented with the additional requirement that $c(t)$ does not simultaneously vanish. This is always the case in Fig.~\ref{fig2}, which is easily checked with the help of the analytical expression for $c(t)$ [Eq.~\eqref{eq: c inf wg}]. 

Further light on the onset of NM effects can be shed by studying in detail
a non-Markovianity measure, which by definition is a function of the \emph{entire} DM (i.e., at all times) \cite{BreuerRMP16}. Out of the many proposed \cite{BreuerRMP16}, we select the geometric measure (GM) \cite{LorenzoPRA13} for its ease of computation and because it facilitates a comparison with the spontaneous emission dynamics in the semi-infinite waveguide where the GM was already used \cite{TufarelliPRA14}. The GM is defined in terms of the DM's determinant
as \cite{LorenzoPRA13}
\begin{equation}
\mathcal N=\!\int\limits_{\partial_t|\det{\bf M}_t|>0}\!\!{\rm d}t\,\frac{{\rm d}}{{\rm d}t} \big|\det{\bf M}_t\big|,\label{def-GM}
\end{equation} 
where the integral is over all the time intervals in which $|\det{\bf M}_t|$ grows in time, and 
\begin{equation}
\det{\bf M}_t= |c(t)|^2 \Delta (t)
\label{eq: det of DM}
\end{equation}
[\cf\eq(\ref{phit1})]. Note that $\mathcal N$ depends on the \emph{modulus} of the  determinant,
which is the volume of the ellipsoid into which the Bloch sphere is transformed by the DM [see \eq(\ref{rt})]. Hence, a non-zero $\mathcal N$ means this volume increases at some time, in contrast to dynamics described by the Lindblad ME in which such an increase cannot occur \cite{LorenzoPRA13}. It is known \cite{BreuerRMP16} that a non-zero GM implies that the dynamics is NM also according to the BLP measure \cite{BreuerPRL09}, which in turn entails NM behavior according to the RHP measure \cite{RivasPRL10}. 

A remarkable property following from \eqs\eqref{def-GM} and \eqref{eq: det of DM} is that if there exists a time such that $\Delta (t)<0$ and $c(t)\neq0$ then ${\big|\det{\bf M}_t\big|}$ must grow at some time. This then brings about that the dynamics is NM according to the GM \eqref{def-GM} and hence NM even according to the BLP and RHP measures. We thus in particular retrieve the sufficient condition for breaking P- and CP-divisibility discussed at the beginning of this section since non-zero BLP (RHP) measure ensures violation of P-divisibility (CP-divisibility) \cite{BreuerRMP16}.
\begin{figure}[tbp]
	\centering
	\includegraphics[scale=1.6]{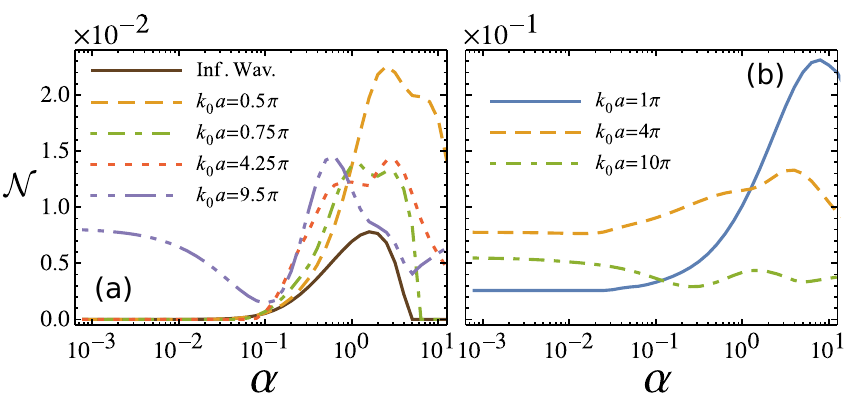}
	\caption{The geometric non-Markovianity measure $\mathcal N$ as a function of the dimensionless wavepacket bandwidth $\alpha$ on a linear-log scale. (a)~Infinite waveguide (solid line) and semi-infinite waveguide for non-integer values of $k_0 a/\pi$. (b)~Semi-infinite waveguide for integer values of $k_0 a/\pi$. Note the different vertical axis scale in the two panels. Non-Markovianity is larger for the semi-infinite-waveguide case (especially for integer $k_0 a/\pi$) because of the time delay in reflecting from the (distant) mirror. ($k=k_0=\omega_0= 20 \Gamma$.)
		\label{fig3}}
\end{figure}

Figure~\figurepanel{fig3}{a} shows $\mathcal N$ for the infinite-waveguide case for a wavepacket carrier frequency resonant with the qubit (solid line). Similarly to the negativity of $\Delta$,
$\mathcal N$ takes significant values only 
around $\alpha\simeq1$ (i.e., $\delta k\simeq\Gamma$), being negligible in particular for quasi-monochromatic wavepackets. The values of $\alpha$ yielding $\mathcal N\neq0$ are also such that the dynamics is NM according to the BLP measure \cite{BreuerPRL09} and even the RHP measure \cite{RivasPRL10}, the latter meaning that rates $\gamma_{\pm,z}(t)$ in the ME (\ref{ME}) break the condition of being positive at all time.

The behavior of $\mathcal N$ changes substantially for a semi-infinite waveguide, as shown in \fig\ref{fig3} for several values of the qubit-mirror distance $a$.
First, non-Markovianity is generally larger, even by an order of magnitude in some cases [note the difference in scale between panels (a) and (b)]. Second, $\mathcal N$ can be significant even at $\alpha\simeq0$ (the plane-wave limit), in which case it matches its value in the corresponding spontaneous-emission process
\cite{TufarelliPRA14}. 
For our parameters, the maximum non-Markovianity in this limit occurs near $k_0a=4\pi$. 
Third, $\mathcal N$ exhibits a more structured behavior as a function of $\alpha$,
the shape of $\mathcal N(\alpha)$ being dependent on $k_0 a$ (recall $k_0=\omega_0$). 

The feedback due to the mirror, evident in Eqs.~\eqref{eq:single-excitation delay differential eq appendix} and \eqref{eq:double-excitation delay differential eq full}, generally introduces memory effects 
in the qubit dynamics 
that are expected to cause NM behavior. 
These add to the finite-wavepacket effect already occurring with no mirror, leading generally to enhanced non-Markovianity---note that the semi-infinite-waveguide curves in \fig\ref{fig3} typically lie above the mirrorless one. The non-Markovianity can be especially large when either $2 k_0 a$, the phase corresponding to a qubit-mirror round trip, is an integer multiple of 2$\pi$ and/or the corresponding photon delay time $\uptau=2 a$ is large compared to the atom decay time $\Gamma^{-1}$. In the former case, enhanced NM behavior occurs because a standing wave can form between the mirror and the qubit under these conditions; indeed, it has been shown that a bound state in the continuum exists in this system \cite{TufarelliPRA13}. In the latter case, the fact that the qubit decays completely before the photon returns causes a periodic re-excitation of the qubit---a kind of revival. 

We found numerically that the scattering DM (\ref{phit1}) reduces to that for spontaneous emission \cite{TufarelliPRA14} in the limits of very large and very small $\alpha$, thus in particular explaining the behavior of $\mathcal N$ at $\alpha\simeq0$. In the infinite-waveguide case, this  property can be shown analytically (see Appendix~\ref{sec: 3 functions}). Physically, these limits can be viewed as follows. 
When $\alpha\simeq0$, the wavepacket is so spread out spatially that the photon density at the qubit is negligible: the qubit effectively sees a vacuum, hence behaving as in spontaneous emission. This clarifies why NM effects cannot occur without the mirror for a quasi-plane-wave (the emission DM in an infinite waveguide is clearly Markovian). When $\alpha\gg1$ in contrast, the photon is very localized at the qubit position. The energy-time uncertainly principle then implies that the photon passes too fast for the qubit to sense, hence the qubit again behaves as if the field were in the vacuum state.

Since non-Markovianity measures are generally not equivalent \cite{BreuerRMP16} it is natural to wonder whether the outcomes of our analysis in Fig.~\ref{fig3} for the GM hold qualitatively if a different measure is used, for instance the widely adopted BLP measure \cite{BreuerPRL09}. While a comprehensive comparative study of different measures is beyond the scope of the present paper, we computed the BLP measure ${\cal N}_{\rm BLP}$ for some representative values of the parameters. For an infinite waveguide, the behavior of ${\cal N}_{\rm BLP}$ as a function of $\alpha$ is analogous to that of the GM [see Fig.~\figurepanel{fig3}{a}]. In the semi-infinite-waveguide case, ${\cal N}_{\rm BLP}$ overall behaves similarly to the GM but exhibits a less structured shape: for instance, the inflection point for $k_0a=0.5\pi$ in Fig.~\figurepanel{fig3}{a} is absent.

\section{Conclusions}
\label{sec: conclusion}
We studied the open dynamics of a qubit coupled to a 
1D waveguide during single-photon scattering, presenting results for its DM, the corresponding time-dependent ME, and rigorous non-Markovianity 
measures developed in OQS theory. The qubit 
dynamics was shown to have distinctive features that, in particular, do not occur in emission processes. To compute the DM for a semi-infinite waveguide, we solved the scattering time evolution by deriving a spatially non-local delay PDE for the one-photon wavefunction when the qubit is excited.  
For an infinite waveguide, NM behavior occurs when the photon-wavepacket bandwidth $\delta k$ is 
of order the qubit decay rate $\Gamma$. For a semi-infinite waveguide (mirror), time delay effects are an additional source of non-Markovianity, resulting in generally stronger NM effects. 

The system we studied here, a semi-infinite waveguide plus a qubit, is the simplest waveguide QED system with a time delay. Yet the nature and effects of the time delay should be completely generic as there is no fine tuning in our system. We thus expect these main conclusions to also hold in, for instance, the case of two distant qubits coupled to a waveguide which is relevant for long-distance quantum information. 

It is interesting to note [see Eq.~\eqref{eq: det of DM}] that $\Delta(t)$ has the same sign as $\det{\bf M}_t$, hence times can exist at which $\det{\bf M}_t<0$. Among qubit CP maps, those with 
negative determinant are the only ones that break the property of being ``infinitesimally divisible" \cite{WolfCMP08}. This class does not include spontaneous-emission DMs---in particular vacuum Rabi oscillations---where the determinant is always non-negative \cite{LorenzoPRA13}. 
In sharp contrast, the scattering DMs studied here do belong to this class.

Finally, we note that some results here rely solely on the DM structure (\ref{phit1}) that in turn stems solely from having an initial Fock state for the field and the rotating wave approximation. Further investigation of this class of open dynamics is under way \cite{wip}.

\begin{acknowledgments}
We thank I.-C.\ Hoi, S.\ Lorenzo and B.\ Vacchini for invaluable discussions. We acknowledge financial support from U.S.\ NSF (Grant No.\ PHY-14-04125) and the Fulbright Research Scholar Program. 
\end{acknowledgments}

\appendix

\section{Derivation of the time-dependent ME}
\label{sec: MEcalc}
In this Appendix, we present some details of the derivation of the time-dependent ME, Eq.\,\eqref{ME}. In particular, we express both Eqs.\,(\ref{ME}) and (\ref{L2}) using as a basis the four Hermitian operators $\{\hat G_i\}= \{\openone/\sqrt{2},\hat \sigma_x/\sqrt{2},\hat \sigma_y/\sqrt{2},\hat \sigma_z/\sqrt{2}\}$ with $i=0,1,2,3$, respectively; recall that $\Tr\{\hat G_i\hat G_j\}= \delta_{ij}$. An operator $\rho$ (and so in particular a density operator) can be decomposed as $\rho= \sum_{i=0}^3 r_i \hat G_i$ with $r_i=\Tr\{\hat G_i\rho\}$, hence the 4-dimensional \emph{real} vector $\bf r$ is a representation of the density operator $\rho$. 
A map is analogously represented by a $4{\times}4$ transformation matrix. 

For ${\cal L}_t$ we start by noting that the dynamical map can be expressed as
\begin{align}
\Phi _t[\rho]&=\sum_{j} \Phi _t(r_j\hat G_j)
=\sum_{j} r_j \Phi _t\hat G_j
=\sum_{j} r_j \sum_k \Tr\left\{\hat G_k\Phi _t \hat G_j\right\}\hat G_k\nonumber\\
&= \sum_k \left(\sum_j\Tr\{\hat G_k\Phi _t\hat G_j\}r_j\right)\hat G_k
=\sum_k\left(\mathbf F \, \mathbf r\right)_k\hat G_k,
\end{align}
where we used the linearity of $\Phi_t$ and defined the entries of the $4{\times} 4$ matrix $\mathbf F$ as
\begin{equation}\label{Fmat}
F_{kj}=\Tr \left\{\hat G_k \Phi_t\hat G_j\right\}.
\end{equation}
The matrix $\mathbf F$ thus represents the map $\Phi_t$ (we drop the time dependance for simplicity). The composition of two maps is correspondingly turned into the matrix product of the associated matrices [each defined analogously to \eq(\ref{Fmat})]. Hence, if $\mathbf L$ is the 4$\times4$ matrix associated with map $\mathcal L_t$ [see \eq(\ref{L2})], it is given by
\begin{equation}
\mathbf L=\dot{\mathbf F}\,\mathbf F^{-1}.
\label{FF}
\end{equation}
We are thus led to compute the (time-dependent) matrix $\bf F$, calculate its derivative $\dot {\bf F}$ and inverse $\bf F^{-1}$, and finally take the matrix product (\ref{FF}). To calculate $\bf F$ we use \eqs(\ref{phit1}) and (\ref{Fmat}), where the matrix elements $\rho_{ij}$ entering \eq(\ref{phit1}) are now the matrix entries of operators $\{\hat G_j\}$ (for instance, $\hat G_1=\hat\sigma_x/\sqrt{2}$ has entries $(G_1)_{ee}= (G_1)_{gg}=0$ and $(G_1)_{eg}= (G_1)_{ge}=1/\sqrt{2}$). By proceeding in this way, matrix $\bf F$ reads
\begin{equation}\label{Ft}
{\bf F}=\left(\begin{array}{cccc}
1&0&0&0\\
0&\re[c(t)]&\im[c(t)]&0\\
0&-\im[c(t)]&\re[c(t)]&0\\
p_e(t)+ p_g(t)-1&0&0& p_e(t)- p_g(t)
\end{array}\right).
\end{equation}
Using this and \eq(\ref{FF}), we find that matrix ${\bf L}$ is 
\begin{equation}\label{Lt}
{\bf L}=\left(
\begin{array}{cccc}
0 & 0 & 0 & 0 \\
0 & \re\left[\frac{\dot c (t)}{c(t)}\right]& \im\left[\frac{\dot c (t)}{c(t)}\right]& 0 \\
0 & -\im\left[\frac{\dot c (t)}{c(t)}\right]  &\re\left[\frac{\dot c (t)}{c(t)}\right] & 0 \\
\frac{\dot p_e(t) + \dot p_e(t)}{p_g(t) + p_e(t)-1 }+2\,\frac{\, \left[1- p_g(t)\right]\dot p_e(t) + p_e(t)  \dot p_g(t)}{1- p_g(t) - p_e(t)}
& 0 & 0 & \frac{\dot p_g(t) + \dot p_e(t)}{p_g(t) + p_e(t)-1 }\
\end{array}
\right).
\end{equation}
This shows that \eq(\ref{Lrho}) holds with the generator $\mathcal L_t$ whose 4$\times$4-matrix representation is given by \eq(\ref{Lt}). 

The remaining step is to show that the generator can indeed be expressed as the right-hand side of (\ref{ME}). To this aim, we consider \eq (\ref{ME}) without specifying $\hat H(t)$, $\gamma_{\pm}(t)$ and $\gamma_z(t)$, work out its 4$\times$4-matrix representation, impose that it yields $\dot{\bf r}= {\bf L}{\bf r}$ with ${\bf L}$ given by \eq(\ref{Lt}) and solve for $\hat H(t)$, $\gamma_{\pm}(t)$ and $\gamma_z(t)$. 
Thus let us define
\begin{equation}
\tilde{\mathcal L}_t [\rho]=\!-i \left[\frac{S(t)}{2}\,\hat\sigma_+\hat \sigma_-+ \mu(t)\hat\sigma_++ \mu^*(t)\hat\sigma_- ,\rho\right]+\! \gamma_-(t)\mathcal L_-[\rho]+\! \gamma_+(t)\mathcal L_+[\rho]+\!\gamma_z(t)\mathcal L_z[\rho]\label{ME2}
\end{equation}
and call $\tilde{\bf L}$ the associated 4$\times4$ matrix. To compute $\tilde{\bf L}$, in \eq(\ref{ME2}) we replace $\rho=\sum_i r_i\hat G_i$ and calculate $\Tr\{\hat A\, \hat G_i\,\hat B\}$ with $\hat A,\hat B=\hat G_0,...,\hat G_3$, obtaining 
\begin{equation}\label{Lt-ansatz}
{\bf \tilde L}=\left(
\begin{array}{cccc}
0 & 0 & 0 & 0 \\
0 &- \frac{\gamma_+(t)+ \gamma_-(t)}{2}- 2{\gamma_z(t)}&-  \frac{S(t)}{2} & - 2\,\im[\mu(t)]\\
0 & \frac{S(t)}{2}  &- \frac{\gamma_+(t)+ \gamma_-(t)}{2}- 2{\gamma_z(t)}& - 2\,\re[\mu(t)] \\
\gamma_+(t)- \gamma_- (t) & 2\,\im[\mu(t)] &2\,\re[\mu(t)] & -\left[\gamma_+(t)+ \gamma_-(t)\right] \\
\end{array}
\right).
\end{equation}
We next require that \eq(\ref{Lt-ansatz}) equals \eq(\ref{Lt}). Upon comparison of these two equations, we immediately get
$\mu(t)=0$, while $S(t)=-2\im\left[\dot c(t)/ c(t)\right]$. Moreover, by requiring the entries $L_{22}$, $ L_{44}$ and $L_{41}$ of matrix (\ref{Lt-ansatz}) to match the corresponding ones of (\ref{Lt}), we find the three rates given in Eq.\,\eqref{eq: time-dep decay rates}.

\section{Calculations for the infinite-waveguide case}
\label{sec: inf wg calculation}

Here, we present details of the calculation of the time-dependent wavefunctions in both the one- and two-excitation sectors that are needed for the explicit calculation of the dynamical map \eq\eqref{phit1} in  the infinite-waveguide case. Following the main text, we refer to $\varphi(x)$ as a single-photon exponential wavepacket 
of the form \eq\eqref{eq: exponential wavepacket}.
Further technical details, including the study of other possible initial conditions, are given in 
the Supplementary Material \cite{SupMat}. 

\subsection{One-excitation sector}

This is the scattering process corresponding to \eq (\ref{ugphi}) in the infinite-waveguide case,
based on which 
the ansatz for the time-dependent wavefunction reads
\begin{equation}
|\Psi_1(t)\rangle=\int dx\left[\phi_\text{R}(x, t)\hat{a}_\text{R}^\dagger(x)+\phi_\text{L}(x, t)\hat{a}_\text{L}^\dagger(x)\right]|0\rangle+e(t)\hat\sigma_+|0\rangle,
\label{eq:single-excitation wavefunction}
\end{equation}
where $\phi_\text{R/L}(x, t)$ is the wavefunction of the right-/left-going photon.

Imposing the Schr\"{o}dinger equation, $i\partial_t|\Psi_1(t)\rangle=\hat{H}|\Psi_1(t)\rangle$, yields the three coupled equations
\begin{subequations}
	\begin{align}
	i\partial_t\phi_\text{R}(x,t)&=-i\partial_x\phi_\text{R}(x,t)+V e(t)\delta(x),\\
	i\partial_t\phi_\text{L}(x,t)&=i\partial_x\phi_\text{L}(x,t)+V e(t)\delta(x),\\
	i\frac{d}{dt}e(t)&=\omega_0 e(t)+V\left[\phi_\text{R}(0, t)+\phi_\text{L}(0, t)\right]
	\label{eq:2LS wavefunction}.
	\end{align}
	\label{eq:single-excitation differential equation infinite}
\end{subequations}
The equations for $\phi_\text{R/L}(x,t)$ can be formally integrated by Fourier transform, yielding
\begin{subequations}
	\label{eq:one-photon formal solution}
\begin{align}
\phi_\text{R}(x,t)&=\phi_\text{R}(x-t, 0)-i V e(t-x)\theta(x)\theta(t-x),\\
\phi_\text{L}(x,t)&=\phi_\text{L}(x+t, 0)-i V e(t+x)\theta(-x)\theta(t+x),
\end{align}
\end{subequations}
where we set $\theta(0)\equiv1/2$. The first term on each righthand side describes the free-field behavior, while the second one can be interpreted as a source term originating from qubit emission at an earlier time. Note that causality is preserved as it should be. Eqs.~\eqref{eq:one-photon formal solution} immediately entail $\phi_\text{R}(0,t)+\phi_\text{L}(0,t)=\phi_\text{R}(-t,0)+\phi_\text{L}(t,0)-iVe(t)$,
which once substituted in Eq.~\eqref{eq:2LS wavefunction} yields a time-local first-order differential equation for $e(t)$
\begin{equation}
\frac{d}{dt}e(t)=-\left(i\omega_0 +\frac{\Gamma}{2}\right) e(t)-i V \left[\phi_\text{R}(-t,0)+\phi_\text{L}(t,0)\right].
\label{eq:2LS differential equation}
\end{equation}
Imposing the initial conditions $\phi_\text{R}(x, 0)=\varphi(x)$, $\phi_\text{L}(x, 0)=e(0)=0$, we obtain,
\begin{equation}
e(t)=\frac{i \sqrt{\alpha  \Gamma ^2/2} \left(e^{-(i k+\alpha  \Gamma/2) t}-e^{-(i\omega_0+\Gamma  /2)t}\right)}{k-\omega_0+i\Gamma (1-\alpha) /2 } .
\label{et}
\end{equation}
By using \eq(\ref{et}) in Eqs.~\eqref{eq:one-photon formal solution}, one then obtains the photon wavefunctions $\phi_\text{R/L}(x,t)$.

\subsection{Two-excitation sector}

Based on \eq(\ref{uephi}), the ansatz for the time-dependent wavefunction reads
\begin{align}
|\Psi_2(t)\rangle&=\int \!\!dx
\left[\psi_\text{R}(x,t)\hat{a}_\text{R}^\dagger(x)+\psi_\text{L}(x,t)\hat{a}_\text{L}^\dagger(x) \right]\hat{\sigma}_+|0\rangle
+\iint \!\!dx_1 dx_2\Bigl[\chi_\text{RR}(x_1, x_2, t)\frac{\hat{a}_\text{R}^\dagger(x_1)\hat{a}_\text{R}^\dagger(x_2)}{\sqrt{2}}\nonumber\\
&\quad+\chi_\text{RL}(x_1, x_2, t)\hat{a}_\text{R}^\dagger(x_1)\hat{a}_\text{L}^\dagger(x_2)
+\chi_\text{LL}(x_1, x_2,t)\frac{\hat{a}_\text{L}^\dagger(x_1)\hat{a}_\text{L}^\dagger(x_2)}{\sqrt{2}}\Bigr]|0\rangle,
\label{eq:double-excitation wavefunction}
\end{align}
where $\psi_\text{R/L}(x, t)$ is the probability amplitude to have a right-/left-propagating photon at position $x$ with the qubit in the excited state, while $\chi_{\alpha\beta}(x_1, x_2, t)$ is the probability amplitude to have an $\alpha$-propagating photon at position $x_1$ and a $\beta$-propagating photon at position $x_2$ (with the qubit unexcited). Terms $\propto\chi_\text{LR}$ have been incorporated in those $\propto\chi_\text{RL}$ by exploiting the symmetrization property $\chi_\text{LR}(x_1, x_2, t)=\chi_\text{RL}(x_2, x_1, t)$. 
The Schr\"odinger equation then yields five coupled differential equations that read
\begin{subequations}
	\begin{align}
	\partial_t\psi_\text{R}(x,t)&=-\partial_x\psi_\text{R}(x,t)-i\omega_0\psi_\text{R}(x,t)
	-iV\left[\frac{\chi_\text{RR}(0,x,t)+\chi_\text{RR}(x,0,t)}{\sqrt{2}}+\chi_\text{RL}(x,0,t)\right],
	\label{eq:R+2LS}\\
	\partial_t\psi_\text{L}(x,t)&=\partial_x\psi_\text{L}(x,t)-i\omega_0\psi_\text{L}(x,t)
	-iV\left[\frac{\chi_\text{LL}(0,x,t)+\chi_\text{LL}(x,0,t)}{\sqrt{2}}+\chi_\text{RL}(0,x,t)\right]
	\label{eq:L+2LS},\\
	\partial_t\chi_\text{RR}(x_1, x_2, t)&=-\left(\partial_{x_1}+\partial_{x_2}\right)\chi_\text{RR}(x_1, x_2, t)-\frac{iV}{\sqrt{2}}\left[ \psi_\text{R}(x_1, t)\delta(x_2)+\psi_\text{R}(x_2, t)\delta(x_1)\right],
	\label{eq:RR differential equation symmetrized}\\
	\partial_t\chi_\text{RL}(x_1, x_2, t)&=-\left(\partial_{x_1}-\partial_{x_2}\right)\chi_\text{RL}(x_1, x_2, t)-i V \left[\psi_\text{R}(x_1, t)\delta(x_2)+\psi_\text{L}(x_2, t)\delta(x_1)\right],\\
	\partial_t\chi_\text{LL}(x_1, x_2, t)&=\left(\partial_{x_1}+\partial_{x_2}\right)\chi_\text{LL}(x_1, x_2, t)- \frac{iV}{\sqrt{2}}\left[ \psi_\text{L}(x_1, t)\delta(x_2)+\psi_\text{L}(x_2, t)\delta(x_1)\right].
	\label{eq:LL differential equation symmetrized}
	\end{align}
\end{subequations}
Note that the equations for $\chi_\text{RR}$ and $\chi_\text{LL}$ are symmetrized because of the bosonic statistics. Similarly to the previous subsection, we first formally solve for the purely photonic wavefunctions and find
\begin{subequations}
	\begin{align}
	\chi_\text{RR}(x_1, x_2, t)&=\chi_\text{RR}(x_1-t, x_2-t, 0)
	-\frac{iV}{\sqrt{2}}\bigl[\psi_\text{R}(x_1-x_2,t-x_2)\theta(x_2)\theta(t-x_2)\nonumber\\
	&\quad+\psi_\text{R}(x_2-x_1, t-x_1)\theta(x_1)\theta(t-x_1)\bigr],\\
	\chi_\text{RL}(x_1, x_2, t)&=\chi_\text{RL}(x_1-t, x_2+t, 0)
	-iV\bigl[\psi_\text{R}(x_1+x_2, t+x_2)\theta(-x_2)\theta(t+x_2)\nonumber\\
	&\quad+\psi_\text{L}(x_2+x_1, t-x_1)\theta(x_1)\theta(t-x_1)\bigr],\\
	\chi_\text{LL}(x_1, x_2, t)&=\chi_\text{LL}(x_1+t, x_2+t, 0)
	-\frac{iV}{\sqrt{2}}\bigl[\psi_\text{L}(x_1-x_2, t+x_2)\theta(-x_2)\theta(t+x_2)\nonumber\\
	&\quad+\psi_\text{L}(x_2-x_1, t+x_1)\theta(-x_1)\theta(t+x_1)\bigr].
	\end{align}
\end{subequations}
Next, we plug these solutions back into Eqs.~\eqref{eq:R+2LS} and \eqref{eq:L+2LS}, which are those featuring the qubit degree of freedom, under the initial condition that $\chi_{\alpha\beta}(x_1, x_2, 0)=0$ for any $\alpha,\beta={\rm L},{\rm R}$. The resulting pair of equations read
\begin{subequations}
	\label{eq:photon+2LS differential equation}
	\begin{align}
	\partial_t\psi_\text{R}(x,t)&=-\partial_x\psi_\text{R}(x,t)-\left(i\omega_0+\frac{\Gamma}{2}\right)\psi_\text{R}(x,t)
	-\frac{\Gamma}{2}\left[\psi_\text{R}(-x,t-x)+\psi_\text{L}(x,t-x)\right]
	\theta(x)\theta(t-x)
	\label{eq:R+2LS differential equation},\\
	\partial_t\psi_\text{L}(x,t)&=\partial_x\psi_\text{L}(x,t)-\left(i\omega_0+\frac{\Gamma}{2}\right)\psi_\text{L}(x,t)
	-\frac{\Gamma}{2}\left[\psi_\text{R}(x,t+x)+\psi_\text{L}(-x,t+x)\right]
	\theta(-x)\theta(t+x).
	\label{eq:L+2LS differential equation}
	\end{align}
\end{subequations}
These coupled differential equations are non-local with respect to both $x$ and $t$, the non-locality being due to the rightmost ``source terms" that feature the double step functions. Based on the arguments of the step functions, it is natural to partition space-time into the three regions $x\le0$ (${\rm R}_1$), $0<x\le t$ (${\rm R}_2$), and $x>t$ (${\rm R}_3$) in the case of \eq(\ref{eq:R+2LS differential equation}) and $x\le-t$ (${\rm L}_3$), $-t<x\le 0$ (${\rm L}_2$), and $x>0$ (${\rm L}_1$) in the case of \eq(\ref{eq:L+2LS differential equation}), as shown in  \fig\ref{fig:spacetime-stimulated emission}. Then, the differential equations \eqref{eq:photon+2LS differential equation} can be analytically solved in four steps as follows:

\begin{figure}[t]
	\centering
	\includegraphics[scale=0.8]{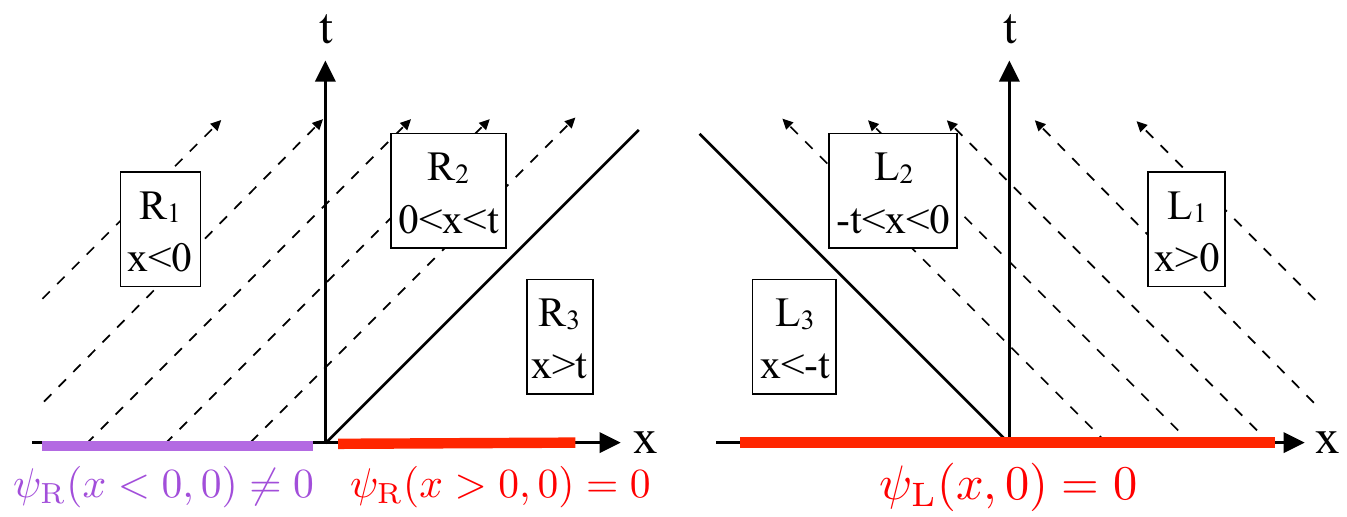}
	\caption{Partitions of space-time in the case of \eqs(\ref{eq:R+2LS differential equation}) and (\ref{eq:L+2LS differential equation}), left and right panels, respectively.}
	\label{fig:spacetime-stimulated emission}
\end{figure}

\begin{enumerate}
	\item[(i)] Solve \eq(\ref{eq:R+2LS differential equation}) for $\psi_\text{R}(x,t)$ in region $\text{R}_1$ under the initial (i.e., boundary) condition $\psi_\text{R}(x,0)=\varphi(x)$. In this region, the source term is identically zero.
	
	\item[(ii)] Solve \eq(\ref{eq:L+2LS differential equation}) for $\psi_\text{L}(x,t)$ in region $\text{L}_1$ under the initial (i.e., boundary) condition $\psi_\text{L}(x,0)=0$. As the source term is also identically zero in this region, we trivially get $\psi_\text{L}(x>0, t)=0$.
	
	\item[(iii)] Solve \eq(\ref{eq:R+2LS differential equation}) for $\psi_\text{R}(x,t)$ in region $\text{R}_2$ under the boundary condition $\psi_\text{R}(0,t)$ [this being fully specified by the solution found at step (i)]. In this region, the source term is non-zero but is fully specified by the solutions $\psi_\text{R}(x<0, t)$ and $\psi_\text{L}(x>0, t)$ worked out at the previous steps (i) and (ii), respectively.
	Note that the initial condition automatically guarantees that the wavefunction is continuous at $x=0$.

	\item[(iv)] Solve \eq(\ref{eq:L+2LS differential equation}) analogously for $\psi_\text{L}(x,t)$ in region $\text{L}_2$ under the boundary condition $\psi_\text{L}(0,t)=0$. In this region, the source term is again fully specified by the solutions $\psi_\text{R}(x<0, t)$ and $\psi_\text{L}(x>0, t)$ obtained in the previous steps.
\end{enumerate}
Finally, $\psi_\text{R}(x,t)$ vanishes identically in region $\text{R}_3$ and, likewise, so does $\psi_\text{L}(x,t)$ in region $\text{L}_3$. This is because causality prevents the wavefunction outside the light cone from being affected by the qubit or input wave. Since initially the wavefunction is zero in this region, it remains so at all times. Hence, the wavefunction is non-zero only in regions R$_1$, R$_2$ and L$_2$.

With the help of \textit{Mathematica} (see Supplementary Material \cite{SupMat}), the above procedure straightforwardly yields analytical expressions for the wavefunctions. 
We checked that, in the steady-state limit $t{\rightarrow}\infty$, the above solution for the wavefunctions in the stimulated-emission problem yields results in full agreement with those obtained via a time-independent approach \cite{RephaeliPRL12}. In particular, the two-photon scattering outcome probabilities $P_\text{RR}$, $P_\text{RL}$ and $P_\text{LL}$ of Ref.~\cite{RephaeliPRL12} are recovered as
\begin{equation}
	P_{\alpha \beta} = \lim_{t \rightarrow \infty} \int_{-\infty}^\infty dx_1 \int_{-\infty}^\infty dx_2 | \chi_{\alpha\beta}(x_1, x_2, t)|^2,
\end{equation}
with $\alpha,\beta \in \{\text{R},\text{L}\}$.

We finally mention that, in the case of an incoming \textit{two-photon} wavepacket (not addressed in the main text), one or more terms $\chi_{\alpha\beta}(x_1, x_2, 0)$ are non-zero and Eqs.~\eqref{eq:photon+2LS differential equation} feature additional terms. For instance, in the case of a left-incoming two-photon wavepacket, the additional term $-i\sqrt{\Gamma/4}\bigl[\chi_\text{RR}(x-t, -t, 0)+\chi_\text{RR}(-t, x-t, 0)\bigr]$ must be added to the right-hand side of \eq(\ref{eq:R+2LS differential equation}). In this case, in the steady-state limit $t\rightarrow\infty$ known results for two-photon scattering (in particular second-order correlation functions) \cite{ShenPRL07,ZhengPRA10,ZhengPRL13,FangPRA15} are recovered, which confirms the effectiveness of our real-space time-dependent approach.


\subsection{Functions $p_{g}(t)$, $p_{e}(t)$ and $c(t)$}
\label{sec: 3 functions}
The three functions (\ref{3functions}), which fully specify the scattering DM (\ref{phit1}), are found from  Eqs.~(\ref{ugphi}), (\ref{uephi}), \eqref{eq:single-excitation wavefunction}, \eqref{eq:double-excitation wavefunction} to be $p_g(t)=|e(t)|^2$,
\begin{align}
p_e(t)&=\Vert\psi_1(t)\Vert^2=\int dx\left[|\psi_\text{R}(x,t)|^2+|\psi_\text{L}(x,t)|^2\right],\\
c(t)&=\langle \phi_1(t)| \psi_1(t)\rangle=\int dx\left[ \phi_\text{R}^*(x,t)\psi_\text{R}(x,t)+\phi_\text{L}^*(x,t)\psi_\text{L}(x,t) \right].
\end{align}
Thus, after using \eqs(\ref{eq:one-photon formal solution}) and (\ref{et}), rescaling time in units of $\Gamma^{-1}$, and setting $k=\omega_0$,
they are explicitly given for the infinite-waveguide case by
\begin{subequations}
\begin{align}
p_g(t)&=2 \alpha  e^{-t} \left[\frac{e^{-\frac{1}{2} (\alpha -1) t}-1}{\alpha -1}\right]^2,
\label{eq: p_g inf wg}\\
p_e(t)&=e^{-(\alpha \!+ \!1)t}\frac{ -4 (\alpha -1)^2+ \left(\alpha ^3-3 \alpha ^2+ \alpha
	+ 5\right) e^{\alpha  t}+ 4 (\alpha -3) \alpha  e^{\frac{1}{2} (\alpha + 1) t}+ 2 \alpha 
	(\alpha + 1) e^t}{(\alpha -1)^2 (\alpha +1)},
\label{eq: p_e inf wg}\\
c(t) &=e^{- (\alpha+ 1 +i
	\omega_0)t}\frac{(\alpha -1)^2 e^{\left(\alpha +\frac{1}{2}\right) t}+4 \alpha 
	e^{\frac{\alpha  t}{2}}-2 (\alpha +1) e^{t/2}}{\alpha ^2-1}.
\label{eq: c inf wg}
\end{align}
\label{eq: p_g p_e c inf wg}
\end{subequations}
From Eqs.~\eqref{eq: p_g inf wg} and \eqref{eq: p_e inf wg}, the quantity $\Delta(t)$ in the infinite-waveguide case is easily obtained as given in \eq(\ref{lambdat2}). 

In the two limits $\alpha\rightarrow0$ and $\alpha\rightarrow\infty$ (see main text), we get
\[
\lim_{\alpha\to 0}p_g(t)=\lim_{\alpha\to \infty}p_g(t)=0,\quad
\lim_{\alpha\to 0}p_e(t)=\lim_{\alpha\to \infty}p_e(t)=e^{-\Gamma t},\quad
\lim_{\alpha\to 0}c(t)=\lim_{\alpha\to \infty}c(t)=e^{-i \omega_0 t}e^{-\frac{\Gamma}{2}t}
\]
with \eq(\ref{phit1}) thereby reducing to
\begin{equation}
\Phi_t[\rho_0]=\left(
\begin{array}{cc}
e^{-\Gamma t}\rho_{ee} & e^{-i \omega_0 t}e^{-\frac{\Gamma}{2}t} \rho_{eg}\\
e^{i \omega_0 t}e^{-\frac{\Gamma}{2}t}\rho_{eg}^*  & \rho_{gg}+(1-e^{-\Gamma t})\rho_{ee}
\end{array}
\right).
\end{equation}
This is the DM of spontaneous emission into an infinite waveguide with a flat spectral density \cite{BreuerBook} obeying the time-independent Lindblad ME $\dot \rho=-i\omega_0 [\hat\sigma_+\hat\sigma_-,\rho]+({\Gamma}/{2})\mathcal L_-[\rho]$, which is thus manifestly Markovian.

\subsection{Study of function $\Delta (t)$}
\label{sec: analysis of Delta}
From Eqs.~\eqref{eq: p_g inf wg} and \eqref{eq: p_e inf wg}, the quantity $\Delta(t)$ in the infinite-waveguide case is easily obtained as in \eq(\ref{lambdat2}). This is such that $\Delta(0)=1$ and $\Delta(t\rightarrow\infty)=0$. We will prove that, based on the analytic function (\ref{lambdat2}), $\Delta (t)$ has a \emph{single} stationary point for $\alpha\!\le 5$ and \emph{no} stationary points for $\alpha>5$. 

The time derivative of function (\ref{lambdat2}) is calculated as
\begin{equation}
\frac{{\rm d}\Delta }{{\rm d}t}=e^{-(\alpha +1) t}\,\,\frac{4 (\alpha -1)+(5-\alpha) e^{\alpha  t}-4
	\alpha  \,e^{\frac{\alpha +1}{2} t}}{\alpha -1}\!=e^{-(\alpha +1) t}\,\,\frac{f(t)-g(t)}{\alpha -1}\,\label{lambdader}
\end{equation}
with
\begin{equation}
f(t)=5 e^{\alpha  t}+4 \alpha,\,\,\,\,\,g(t)=4 \alpha  e^{\frac{\alpha +1 }{2} t}+\alpha  e^{\alpha  t}+4\,.\label{fg}
\end{equation} 
For $\alpha{\ne}1$, at a stationary point of $\Delta (t)$, thereby, curves $f(t)$ and $g(t)$ cross. Note that the positive functions $f(t)$ and $g(t)$ both monotonically increase with time and so do all their derivatives. Thus there exist either zero or only {one} crossing point, whose occurrence depends on whether $f(t)$ is above or below $g(t)$ at $t=0$ and $t\rightarrow \infty$. A simple calculation yields
$$\frac{f(0)}{g(0)}=\frac{4 \alpha +5}{5 \alpha +4},\,\,\,\,\,\,\lim_{t\to \infty} \,\frac{f(t)}{g(t)}=\left\{\begin{array}{c}
0\,\,\,\,{\rm if}\,\,\,\,\alpha< 1\\
\frac{5}{\alpha}\,\,\,\,{\rm if}\,\,\,\,\alpha> 1
\end{array}\right.\,.$$
By noting that ${f(0)}/{g(0)}> 1$ for $\alpha< 1$ and ${f(0)}/{g(0)}< 1$ for $\alpha> 1$, we see that three cases occur. For $\alpha<1$, $f(t)$ is above $g(t)$ at $t=0$ and below it at $t\rightarrow\infty$, hence a single crossing point occurs. For $1< \alpha\!\le\! 5$, $f(t)$ lies below $g(t)$ at $t=0$ and above it at $t\rightarrow\infty$, hence a single crossing point occurs in this case as well. Finally, for $\alpha>5$, $f(t)$ lies below $g(t)$ both at $t=0$ and at $t\rightarrow\infty$, hence no crossing points occur. 
Function $\Delta (t)$ thereby has a single stationary point for $0\le\alpha\le 5$ and none for $\alpha>5$. One can show that this stationary point is indeed minimum, concluding the proof. Note that we have now included the case $\alpha=1$ since this yields $\Delta(t)=e^{-2 t} \left[e^t (3-2 t)-2\right]$, which exhibits a single stationary point that is a minimum.

\bibliography{WQED_2016,\jobname}

\begin{thebibliography}{86}
\bibitem [{emi()}]{emission}%
  \BibitemOpen
  \href@noop {} {}\bibinfo {note} {By ``spontaneous emission'' or ``emission''
  in short, we refer to the process in which initially the qubit is excited and
  the waveguide is in the vacuum state. Such a process can be
  non-Markovian.}\BibitemShut {Stop}%
\bibitem [{bra()}]{brazil}%
  \BibitemOpen
  \href@noop {} {}\bibinfo {note} {Absorption and dephasing terms like those in
  ME (\ref{ME}) are absent in \rref \cite{ValenteOL16}. This is because
  \cite{ValenteOL16} treats only $\rho_{0}= |g\rangle\langle g|$.}\BibitemShut
  {Stop}%
\bibitem [{not()}]{nota}%
  \BibitemOpen
  \href@noop {} {}\bibinfo {note} {Any infinitesimal quantum map $\mathcal
  M(dt)$ that is completely positive can always be expressed as $\mathcal
  M(dt)=\mathbb{I}+\tilde{\mathcal L} dt$ with $\tilde{\mathcal L}$ being a
  superoperator in Lindblad form with positive rates \cite{BreuerBook}. A
  CP-divisible dynamics is described by a Lindblad ME, where the superoperator
  in Lindblad form with positive rates and the Hamiltonian are in general
  time-dependent \cite{BreuerBook,HallPRA14}}\BibitemShut {NoStop}%
\end{thebibliography}
\end{document}